\input harvmac
\input psfig
\newcount\figno
\figno=0
\def\fig#1#2#3{
\par\begingroup\parindent=0pt\leftskip=1cm\rightskip=1cm\parindent=0pt
\global\advance\figno by 1
\midinsert
\epsfxsize=#3
\centerline{\epsfbox{#2}}
\vskip 12pt
{\bf Fig. \the\figno:} #1\par
\endinsert\endgroup\par
}
\def\figlabel#1{\xdef#1{\the\figno}}
\def\encadremath#1{\vbox{\hrule\hbox{\vrule\kern8pt\vbox{\kern8pt
\hbox{$\displaystyle #1$}\kern8pt}
\kern8pt\vrule}\hrule}}
\def\underarrow#1{\vbox{\ialign{##\crcr$\hfil\displaystyle
 {#1}\hfil$\crcr\noalign{\kern1pt\nointerlineskip}$\longrightarrow$\crcr}}}
% use of underarrow
%A~~~\underarrow{a}~~~B
%
\overfullrule=0pt

\def\HP{{\bf HP}}
\def\WCP{{\bf WCP}}
%macros
%

\def\bar{\overline}
\def\C{{\bf C}}
\def\Z{{\bf Z}}
\def\T{{\bf T}}
\def\S{{\bf S}}
\def\R{{\bf R}}
\def\CP{{\bf CP}}

\font\zfont = cmss10 %scaled \magstep1

\def\bigone{\hbox{1\kern -.23em {\rm l}}}
\def\ZZ{\hbox{\zfont Z\kern-.4emZ}}

\hfill RUHNETC-2001-27
\Title{hep-th/0109152}{\vbox{\centerline{Chiral Fermions from
Manifolds Of $G_2$ Holonomy}}}
\smallskip
\centerline{Bobby Acharya}
\smallskip
\centerline{\it Department of Physics, Rutgers University,
Piscataway, NJ 08854, USA} \smallskip \centerline{and}
\smallskip
\centerline{Edward Witten}
\smallskip
\centerline{\it School of Natural Sciences, Institute for Advanced Study}
\centerline{\it Olden Lane, Princeton, NJ 08540, USA}\bigskip

\medskip

\noindent
%abstract

$M$-theory compactification on a manifold $X$ of $G_2$ holonomy
can give chiral fermions in four dimensions only if $X$ is
singular. A number of examples of conical singularity that give
chiral fermions are known; the present paper is devoted to
describing some additional examples.  In some of them, the physics
can be determined but the metric is not known explicitly, while in
others the metric can be described explicitly but the physics is
more challenging to understand.

 \Date{August, 2001}
%text of paper
\newsec{Introduction}

\def\hat{\widehat}

Compactification of $M$-theory on a seven-manifold  $X$ of $G_2$
holonomy is a natural way to obtain  a four-dimensional theory
with ${\cal N}$ $=1$ supersymmetry. However, in the case of a
smooth manifold $X$, one obtains in this way abelian gauge groups
only without chiral fermions.

Singular manifolds of $G_2$ holonomy offer additional
possibilities.  A variety of different kinds of singularity are
possible in $M$-theory. The $A-D-E$ singularities appear in
codimension four, so they arise in compactification on a K3
surface.  They can be embedded in a manifold of $G_2$ holonomy and
give a mechanism to generate gauge symmetry. In codimension six,
there are many $M$-theory singularities that can appear in
Calabi-Yau threefolds; again, they can be embedded in a manifold
of $G_2$ holonomy.

When a codimension four or codimension six singularity is
embedded in a manifold of $G_2$ holonomy, the resulting physics
largely is determined by what happens on a Calabi-Yau two-fold or
three-fold.
  What is really new for a $G_2$-manifold is the
occurrence of singularities of codimension seven, that is,
isolated singularities (or singularities that are isolated modulo
orbifold singularities).  Since chiral fermions are a phenomenon
that is special to compactification to four dimensions, it is
perhaps not surprising that the isolated singularities turn out
to be important for getting chiral fermions.

The known isolated singularities of $G_2$-manifolds are conical
singularities. A number of examples of isolated conical
singularities of $G_2$-manifolds have been studied recently from
different points of view
 \nref\acharya{B. Acharya, ``On Realising ${\cal N}=1$ Super
Yang-Mills In $M$-Theory,''
hep-th/0011089.}%
\nref\amv{M. F. Atiyah, J. Maldacena, and C. Vafa, ``An $M$-Theory Flop As A Large
$N$ Duality,'' hep-th/0011256.}%
 \nref\aw{M. F. Atiyah and E. Witten,
``$M$-Theory Dynamics On A
Manifold Of $G_2$ Holonomy,'' hep-th/0107177.}%
\nref\csq{M. Cvetic, G. Shiu, and A. M. Uranga, ``Three Family Supersymmetric
Standard-Like Models From Intersecting Brane Worlds,'' hep-th/0107143,
``Chiral Four-Dimensional ${\cal N}=1$ Supersymmetric Type IIA Orientifolds
{}From Intersecting $D6$ Branes,'' hep-th/0107166.}%
\refs{\acharya - \csq}.  Some of them give chiral fermions.
Anomaly cancellation in models with singularities of this kind has
been investigated in \ref\witt{E. Witten, ``Anomaly Cancellation
on Manifolds of $G_2$-holonomy,'' hep-th/0108165.}.

The present paper is devoted to constructing and investigating
two additional classes of examples of conical singularities of
$G_2$-manifolds.  Both classes of example are constructed by
taking the quotient of a conical hyper-Kahler eight-manifold
$\hat X$ by a $U(1)$ symmetry group.  The two classes differ by
what kind of $U(1)$ symmetry group is chosen.

In section two, we use duality with the heterotic string to
motivate one construction, which is quite similar to one described
for Type II strings on a Calabi-Yau threefold in \ref\kv{S. Katz
and C. Vafa, ``Matter From Geometry,'' hep-th/9606086.}. In this
construction, the $U(1)$ group preserves the hyper-Kahler
structure of $\hat X$.
 For these examples, though we predict
the existence of a conical $G_2$ metric, we cannot describe it
explicitly.  This is an interesting open problem. On the other
hand, we can describe explicitly the physics that emerges from
these examples since they are constructed to ``geometrically
engineer'' a desired result. For example, suitably chosen examples
of this type give the usual chiral matter representations for the
familiar grand unified groups such as $SU(5)$, $SO(10)$, and
$E_6$.

\nref\threesas{C. Boyer and K. Galicki, ``3-Sasakian Manifolds,''
hep-th/9810250, in {\it Essays On Einstein Manifolds}, ed. C.
LeBrun and M. Wang (International Press, 1999).}%
 \nref\twist{K. Galicki and H. Lawson, ``Quaternionic Reduction
And
Quaternionic Orbifolds,'' Math. Ann. {\bf 282} (1988) 1.}%
\nref\brysal{R. Bryant and S. Salamon,``Complete Metrics With
Exceptional
Holonomy,'' Duke Math Journal, {\bf 58}  (1989) 829.} %
\nref\gibb{G. Gibbons, D. Page and C. Pope, ``Einstein Metrics on
${\S}^3$, ${\R}^3$ and
$\R^4$-bundles,'' Comm. Math. Phys. {\bf 127} (1990) 529.} %
 In section three, we consider
examples of a different kind, constructed with the same $\hat
X$'s that appear in section two, but dividing by a different kind
of  $U(1)$ symmetry (a subgroup of an $SU(2)$ symmetry that
rotates the complex structures of $\hat X$). Surprisingly, this
also leads to conical metrics of $G_2$ holonomy, metrics that can
be described explicitly using known results on self-dual Einstein
metrics from quaternionic reduction \refs{\threesas,\twist}
together with the construction of $G_2$ metrics on cones over
twistor spaces of self-dual Einstein manifolds
\refs{\brysal,\gibb}.  As the metric in these examples is known,
 the problem is to understand the physics.  This can be done in
 some cases, but in general is not as straightforward as in
 section 2.

\newsec{Singularities From Duality With The Heterotic String}

We start by considering duality with the heterotic string.  The
heterotic string compactified on a Calabi-Yau three-fold $W$ can
readily give chiral fermions. On the other hand, most Calabi-Yau
manifolds participate in mirror symmetry. For $W$ to participate
in mirror symmetry means \ref\syz{A. Strominger, S.-T. Yau, and
E. Zaslow, ``Mirror Symmetry Is $T$ Duality,'' hep-th/9606040,
Nucl. Phys. {\bf B479} (1996) 243.} that, in a suitable limit of
its moduli space, it is a $\T^3$ fibration (with singularities and
monodromies) over a base $Q$. Then, taking the $\T^3$'s to be
small and  using on each fiber the equivalence of the heterotic
string on $\T^3$ with $M$-theory on K3, it follows that the
heterotic string on $W$ is dual to $M$-theory on a seven-manifold
$X$ that is K3-fibered over $Q$ (again with singularities and
monodromies). $X$ depends on the gauge bundle on $W$.  Since the
heterotic string on $W$ is supersymmetric, $M$-theory on $X$ is
likewise supersymmetric, and hence $X$ is a manifold of $G_2$
holonomy.

Since there are many $W$'s that could be used in this
construction (with many possible classes of gauge bundles) it
follows that there are many manifolds of $G_2$ holonomy with
suitable singularities to give nonabelian gauge symmetry with
chiral fermions.  The same conclusion can be reached using
duality with Type IIA, as many six-dimensional Type IIA
orientifolds that give chiral fermions are dual to $M$-theory on
a $G_2$ manifold \csq.

Let us try to use this construction to determine what kind of
singularity $X$ must have.  (The reasoning and the result are very
similar to that given in \kv\ for engineering matter from Type II
singularities.  In \kv, the Dirac equation is derived directly
rather than being motivated -- as we will -- by using duality
with the heterotic string.) Suppose that the heterotic string on
$W$ has an unbroken gauge symmetry $G$, which we will suppose to
be simply-laced (in other words, an $A$, $D$, or $E$ group) and at
level one.  This means that each K3 fiber of $X$ will have a
singularity of type $G$. As one moves around in $X$ one will get
a family of $G$-singularities parameterized by $Q$. If $Q$ is
smooth and the normal space to $Q$ is a smoothly varying family of
$G$-singularities, the low energy theory will be $G$ gauge theory
on $\R^4\times Q$ without chiral multiplets. So chiral fermions
will have to come from singularities of $Q$ or points where $Q$
passes through a worse-than-orbifold singularity of $X$.

We can use the duality with the heterotic string to determine
what kind of singularities are required.  The argument will
probably be easier to follow if we begin with a specific example,
so we will consider the case of the $E_8\times E_8$ heterotic
string with $G=SU(5)$ a subgroup of one of the $E_8$'s.  Such a
model can very easily get chiral ${\bf 5}$'s and ${\bf 10}$'s of
$SU(5)$; we want to see how this comes about, in the region of
moduli space in which $W$ is $\T^3$-fibered over $Q$ with small
fibers, and then we will translate this description to $M$-theory
on $X$.

\subsec{ Computation In Heterotic String Language}

\def\D{{\cal D}}
Let us consider, for example, the ${\bf 5}$'s.  The commutant of
$SU(5)$ in $E_8$ is a second copy of $SU(5)$, which we will
denote as $SU(5)'$.  Since $SU(5)$ is unbroken, the structure
group of the gauge bundle $E$ of $W$ reduces from $E_8$ to
$SU(5)'$. Massless fermions in the heterotic string transform in the adjoint
representation of $E_8$.  The part of the adjoint representation of $E_8$ that
transforms as ${\bf 5}$ under $SU(5)$ transforms as ${\bf 10}$
under $SU(5)'$. So to get massless chiral ${\bf 5}$'s of $SU(5)$,
we must look at the Dirac equation $\D$ on $W$ with values in the
${\bf 10}$ of $SU(5)'$; the zero modes of that Dirac equation
will give us the massless ${\bf 5}$'s of the unbroken $SU(5)$.

We denote the generic radius of the $\T^3$ fibers as $\alpha$,
and we suppose that $\alpha$ is much less than the characteristic
radius of $Q$.  This is the regime of validity of the argument for
duality with $M$-theory on $X$ (and the analysis of mirror
symmetry \syz). For small $\alpha$,
 we can solve the Dirac equation on $X$ by first solving it along the
fiber, and then along the base. In other words, we write
$\D=\D'+\D''$, where $\D''$ is the Dirac operator along the fiber
and $\D'$ is the Dirac operator along the base. The eigenvalue of
$\D''$ gives an effective ``mass'' term in the Dirac equation on
$Q$.  For generic fibers of $W\to Q$, as we explain momentarily,
the eigenvalues of $\D''$ are all nonzero and  of order
$1/\alpha$. This is much too large to be canceled by the behavior
of $\D'$. So zero modes of $\D$ are localized near points in $Q$
above which $\D''$ has a zero mode.

When restricted to a $\T^3$ fiber, the $SU(5)'$ bundle $E$ can be
described as a flat bundle with monodromies around the three
directions in $\T^3$.  For generic monodromies, every vector in
the ${\bf 10}$ of $SU(5)'$ undergoes non-trivial ``twists'' in
going around some (or all) of the three directions in $\T^3$.
When this is the case, the minimum eigenvalue of $\D''$ is of
order $1/\alpha$.  A zero mode of $\D''$ above some point $P\in
Q$  arises precisely if for some vector  in the ${\bf 10}$, the
monodromies in the fiber are all trivial.

This means that the monodromies lie in the subgroup of $SU(5)'$
that leaves fixed that vector.  If we represent the ${\bf 10}$ by
an antisymmetric $5\times 5$ matrix $B^{ij}$, $i,j=1,\dots,5$,
then the monodromy-invariant vector corresponds to an
antisymmetric matrix $B$ that has some nonzero matrix element,
say $B^{12}$; the subgroup of $SU(5)'$ that leaves $B$ invariant
is clearly then a subgroup of $SU(2)\times SU(3)$ (where in these
coordinates, $SU(2)$ acts on $i,j=1,2$ and $SU(3)$ on
$i,j=3,4,5$).  Let us consider the case (which we will soon show
to be generic)  that $B^{12}$ is the only nonzero matrix element
of $B$.  If so, the subgroup of $SU(5)'$ that leaves $B$ fixed is
precisely $SU(2)\times SU(3)$. There is actually a distinguished
basis in this problem -- the one that diagonalizes the
monodromies near $P$ -- and it is in this basis that $B$ has only
one nonzero matrix element.

The commutant of $SU(2)\times SU(3)$ in $E_8$ is $SU(6)$.  So over
the point $P$, the monodromies commute not just with $SU(5)$ but
with $SU(6)$.  Everything of interest will happen inside this
$SU(6)$. The reason for this is that the monodromies at $P$ give
large masses to all $E_8$ modes except those in the adjoint of
$SU(6)$.   So we will formulate the rest of the discussion as if
the heterotic string gauge group were just $SU(6)$, rather than
$E_8$.  Away from $P$, the monodromies break $SU(6)$ to
$SU(5)\times U(1)$ (the global structure is actually $U(5)$).
Restricting the discussion from $E_8$ to $SU(6)$ will mean
treating the vacuum gauge bundle as a $U(1)$ bundle (the $U(1)$
being the second factor in $SU(5)\times U(1)\subset SU(6)$)
rather than an $SU(5)'$ bundle.

The fact that, over $P$, the heterotic string has unbroken
$SU(6)$ means that, in the $M$-theory description, the fiber over
$P$ has an $SU(6)$ singularity.  Likewise, the fact that away from
$P$, the heterotic string has only $SU(5)\times U(1)$ unbroken
means that the generic fiber, in the $M$-theory description, must
contain an $SU(5)$ singularity only, rather than an $SU(6)$
singularity. As for the unbroken $U(1)$, in the $M$-theory
description it must be carried by the $C$-field.

If we move away from the point $P$ in the base, the vector $B$ in
the ${\bf 10}$ of $SU(5)'$ is no longer invariant under the
monodromies.  Under parallel transport around the three
directions in $\T^3$, it is transformed by phases $e^{2\pi
i\theta_j}$, $j=1,2,3$.  Thus, the three $\theta_j$ must all
vanish to make $B$ invariant. As $Q$ is three-dimensional, we
should expect generically that the point $P$ above which the
monodromies are trivial is isolated. (Now we can see why it is
natural to consider the case that, in the basis given by the
monodromies near $P$, only one matrix element of $B$ is nonzero.
Otherwise, the monodromies could act separately on the different
matrix elements, and it would be necessary to adjust more than
three parameters to make $B$ invariant.  This would be  a less
generic situation.) We will only consider the (presumably
generic) case that $P$ is disjoint from the singularities of the
fibration $W\to Q$. Thus, the $\T^3$ fiber over $P$ is smooth (as
we have implicitly assumed in introducing the monodromies on
$\T^3$).

Since the fermion zero mode we are looking for is localized near
$P$, the global behavior of $Q$ does not matter, and we can
replace $Q$ by $\R^3$ and take  $P$ to be the origin in $\R^3$.
Also, since we are working in a small neighborhood of $P$, a
variation in the fibers of $W$ will not be important and we can
take the fibration to be a simple product. So we can replace $W$
by $\R^3\times \T^3$.

The generic behavior near $P$ is that the monodromy angles
$\theta_j$ have a nondegenerate common zero, in which case we can
use the $\theta_j$ as local coordinates on $\R^3$ near $P$.  The
chirality that we will get by solving the Dirac equation depends
only on the local behavior of the $\theta_j$ near $P$.  We could
determine the answer from a general index theorem, but instead we
will solve the Dirac equation in a special case.  In $W=\R^3\times
\T^3$, we take $\R^3$ to have linear coordinates $x_j$,
$j=1,2,3$, and $\T^3$ to have angular coordinates $\phi_j$,
$j=1,2,3$, with the complex structure on $W$ such that $z_j=
x_j+i\phi_j$ is holomorphic. Thus, holomorphically,
$W={\C}^*{\times} {\C}^*{\times} {\C}^*$ with $z_j$ a holomorphic
function on the $j^{th}$ copy of $\C^*$. We assume that the
Kahler form on $W$ is $\omega=\alpha^2\sum_j dx_j\wedge d\phi_j$;
this determines also the Kahler metric. We suppose that the
monodromy angles are $\psi_j= \sigma_j f(x_j)$, where $\sigma_j$
are small nonzero real constants  (with a further restriction
discussed momentarily) and $f(x)$ is a monotonic function that is
$-1$ for $x\to -\infty$, 0 at $x=0$, and $1$ for $x\to\infty$.
  We can get these monodromies with an abelian gauge field
$A=\sum_j \sigma_j f(x_j)d\phi_j$. The curvature of the bundle is
then $F=dA=\sum_j\sigma_j f'(x_j)dx^j d\phi_j$.

Next, we would like to impose the condition that the bundle is
stable, and the connection $A$ obeys the hermitian Yang-Mills
equation $\omega\wedge\omega \wedge F=0$ ($\omega$ being the
Kahler form).  With our assumed Kahler form, $\omega\wedge
\omega\wedge F$ is a multiple of $\sum_j\sigma_jf'(x_j)$.  To obey
the hermitian Yang-Mills equation, we need to make a complex
gauge transformation to set $\omega\wedge\omega\wedge F$ to zero.
This can be done in a unique way if and only if $\int
\omega\wedge\omega\wedge F=0$, which in our case (as  $f'(x_j)>0$)
is so precisely if two of the $\sigma_j$ are positive and one is
negative, or vice-versa.  The Dirac equation that we consider
next is invariant under a complex gauge transformation, so we
need not consider the details of modifying $A$ to obey the
hermitian Yang-Mills equation.

It is easy to solve the Dirac equation in this situation.  We have
$\D=\sum_j \D_j$, where $\D_j$ is a Dirac equation on the
$j^{th}$ copy of $\C^*$.  The $\D_j$ anticommute, and a zero mode
of $\D$ must be annihilated by each of the $\D_j$.  We can
explicitly write $\alpha \D_j$ as \eqn\poko{\left(\matrix{ 0 &
{\partial\over \partial x_j}+i\left({\partial\over\partial
\phi_j}+i\sigma_jf(x_j)\right) \cr {\partial\over \partial
x_j}-i\left({\partial\over\partial \phi_j}+i\sigma_jf(x_j)\right)
& 0 \cr}\right).}  In this basis, the upper and lower components
correspond to positive and negative chirality, respectively, or --
upon making the standard relation of spinors on a Calabi-Yau
manifold to $(0,q)$-forms -- to $(0,0)$ and $(0,1)$-forms,
respectively.  If the $\sigma_j$ are small, zero
modes are independent of $\phi_j$.  The possible zero modes are
\eqn\ruggo{\left(\matrix{\exp(-\sigma_j\int_{0}^{x_j}dt f(t))
\cr\cr 0 \cr}\right) ~~{\rm and} ~~ \left(\matrix{0 \cr \cr
\exp(\sigma_j\int_{0}^{x_j}dt f(t)) \cr }\right).} The first of
these is normalizable if $\sigma_j>0$ and the second is
normalizable if $\sigma_j<0$.

\def\hat{\widehat}
After tensoring together the zero modes of the $\D_j$ for
$j=1,2,3$, we learn that there is always precisely one zero mode
in the ${\bf 5}$ of $SU(5)$.  It is a $(0,1)$-form or
$(0,2)$-form depending on whether the number of negative
$\sigma_j$ is 1 or 2.  Thus, the chirality of this zero mode
depends on the sign of the product $\sigma_1\sigma_2\sigma_3$.
Since $F\wedge F\wedge F$, whose integral gives the third Chern
class of the gauge bundle, is proportional in our example to
$\sigma_1\sigma_2\sigma_3$, this is a local version of the
familiar fact that the chiral asymmetry in Calabi-Yau
compactification (the number of left-handed ${\bf 5}$'s minus the
number of right-handed ${\bf 5}$'s) is equal to the integral of
the third Chern class.\foot{Since $W=\C^*\times \C^*\times \C^*$ is
noncompact in our model calculation, an index theorem governing
this situation would involve not only the integral of $F\wedge
F\wedge F$ but also the behavior of the gauge field at
infinity.}  Of course, if we want fermion zero modes transforming
as $\overline {\bf 5}$, we get them by complex conjugation, which
reverses the chirality; so if there is a left-handed ${\bf 5}$,
there is a right-handed $\bar{\bf 5}$, and vice-versa.

\subsec{ Description In $M$-Theory Language}

So we have found a local structure in the heterotic string that
gives a net chirality -- the number of massless left-handed $
{\bf 5}$'s minus right-handed ${\bf 5}$'s -- of one. Let us see in
more detail what it corresponds to in terms of $M$-theory on a
manifold of $G_2$ holonomy.

Here it may help to review the case considered in \kv, where the
goal was geometric engineering of charged matter on a Calabi-Yau
threefold in Type IIA.  What was considered there was a Calabi-Yau
three-fold $R$, fibered by K3's with a base $Q'$, such that over
a distinguished point $P\in Q'$ there is a singularity of type
$\hat G$, and over the generic point in $Q'$ this singularity is
replaced by one of type $G$ -- the rank of $\hat G$ being one
greater than that of $G$.  In our example, $\hat G=SU(6)$ and
$G=SU(5)$. In the application to Type IIA, although $R$ also has a
Kahler metric, the focus is on the complex structure.  For $\hat
G=SU(6)$, $G=SU(5)$, let  us describe the complex structure of
$R$  near the singularities. The $SU(6)$ singularity is described
by an equation $xy=z^6$. Its ``unfolding'' depends on five
complex parameters and can be written $zy=z^6 +P_4(z)$, where
$P_4(z)$ is a quartic polynomial in $z$. If -- as in the present
problem -- we want to deform the $SU(6)$ singularity while
maintaining an $SU(5)$ singularity, then we must pick $P_4$ so
that the polynomial $z^6+P_4$ has a fifth order root.  This
determines the deformation to be
\eqn\okok{xy=(z+5\epsilon)(z-\epsilon)^5,} where we interpret
$\epsilon$ as a complex parameter on the base $Q'$.  Thus, \okok\
gives the complex structure of the total space $R$.

What is described in \okok\ is the partial unfolding of the
$SU(6)$ singularity, keeping an $SU(5)$ singularity.  In our
$G_2$ problem, we need a similar construction, but we must view
the $SU(6)$ singularity as a hyper-Kahler manifold, not just a
complex manifold.  In unfolding the $SU(6)$ singularity as a
hyper-Kahler manifold,  each complex parameter in $P_4$ is
accompanied by a real parameter that controls the area of an
exceptional divisor in the resolution/deformation of the
singularity.  The parameters are thus not five complex parameters
but five triplets of real parameters. (There is an $SU(2)$
symmetry that rotates each triplet.  In perturbative string
theory, each triplet combines with a theta angle to make up a
four-component hypermultiplet.)

To get a $G_2$-manifold, we must combine the complex parameter
seen in \okok\ with a corresponding real parameter.  Altogether,
this will give a three-parameter family of deformations of the
$SU(6)$ singularity (understood as a hyper-Kahler manifold) to a
hyper-Kahler manifold with an $SU(5)$ singularity.  The parameter
space of this deformation is what we have called $Q$, and the
total space is a seven-manifold that is our desired singular
$G_2$-manifold $X$, with a singularity that produces the chiral
fermions that we analyzed above in the heterotic string language.

\def\H{{\bf H}}
To find the hyper-Kahler unfolding of the $SU(6)$ singularity
that preserves an $SU(5)$ singularity is not difficult, using
Kronheimer's description of the general unfolding via a
hyper-Kahler quotient \ref\kron{P. Kronheimer, ``The Construction
of ALE Spaces As Hyper-Kahler Quotients,'' J. Diff. Geom. {\bf
28} (1989) 665, ``A Torelli-Type Theorem For Gravitational
Instantons,'' J. Diff. Geom. {\bf 29} 685. }. At this stage, we
might as well generalize to $SU(N)$, so we consider a hyper-Kahler
unfolding of the $SU(N+1)$ singularity to give an $SU(N)$
singularity.  The unfolding of the $SU(N+1)$ singularity is
obtained by taking a system of $N+1$ hypermultiplets
$\Phi_0,\Phi_1,\dots \Phi_N$ with an action of $K=U(1)^N$.  Under
the $i^{th}$ $U(1)$ for $i=1,\dots, N$, $\Phi_i$ has charge $1$,
$\Phi_{i-1}$ has charge $-1$, and the others are neutral. This
configuration of hypermultiplets and gauge fields is known as the
quiver diagram of $SU(N+1)$ and appears in studying $D$-branes
near the $SU(N+1)$ singularity \ref\dm{M. R. Douglas and G. Moore,
``$D$-Branes, Quivers, and ALE Instantons,'' hep-th/9603167.}. We
let $\H$ denote $\R^4$, so the hypermultiplets parameterize
$\H^{N+1}$, the product of $N+1$ copies of $\R^4$. The
hyper-Kahler quotient of ${\bf H}^{N+1}$ by $K$ is obtained by
setting the $\vec D$-field (or components of the hyper-Kahler
moment map) to zero and dividing by $K$.  It is denoted
$\H^{N+1}//K$, and is isomorphic to the $SU(N+1)$ singularity
$\R^4/\Z_{N+1}$. Its unfolding is described by setting the $\vec
D$-fields equal to arbitrary constants, not necessarily zero. In
all, there are $3N$ parameters in this unfolding -- three times
the dimension of $K$ -- since for each $U(1)$, $\vec D$ has three
components, rotated by an $SU(2)$ group of $R$-symmetries.

\def\hat{\widehat}

We want a partial unfolding keeping an $SU(N)$ singularity.  To
describe this, we keep $3(N-1)$ of the parameters equal to zero
and let only the remaining three vary; these three will be simply
the values of $\vec D$ for one of the $U(1)$'s.  To carry out this
procedure, we first write $K=K'\times U(1)'$ (where $U(1)'$
denotes a chosen $U(1)$ factor of $K=U(1)^N$).  Then we take the
hyper-Kahler quotient of $\H^{N+1}$ by $K'$ to get a hyper-Kahler
eight-manifold  $\hat X=\H^{N+1}//K'$, after which we take the
{\it ordinary} quotient, not the hyper-Kahler quotient, by
$U(1)'$ to get a seven-manifold $X=\hat X/U(1)'$ that should
admit a metric of $G_2$-holonomy. $X$ has a natural map to
$Q=\R^3$ given by the value of the $\vec D$-field of $U(1)'$ --
which was not set to zero -- and this map gives the fibration of
$X$ by hyper-Kahler manifolds.

In the present example, we can easily make this explicit.  We take
$U(1)'$ to be the ``last'' $U(1)$ in $K=U(1)^N$, so $U(1)'$ only
acts on $\Phi_{N-1}$ and $\Phi_N$.  $K'$ is therefore the product
of the first $N-1$ $U(1)$'s; it acts trivially on $\Phi_N$, and
acts on $\Phi_0,\dots,\Phi_{N-1}$ according to the standard quiver
diagram of $SU(N)$.  So the hyper-Kahler quotient $\H^{N+1}//K'$
is just $(\H^N//K')\times \H'$, where $\H^N//K'$ is the $SU(N)$
singularity, isomorphic to $\H/\Z_N$, and $\H'$ is parameterized
by $\Phi_N$. So finally, $X$ will be $(\H/\Z_N\times \H')/U(1)'$.
To make this completely explicit, we just need to identify the
group actions on $\H$ and $\H'$. If we parameterize $\H$ and
$\H'$ respectively by pairs of complex variables
$\left(\matrix{a\cr b\cr}\right)$ and $\left(\matrix{a' \cr
b'\cr}\right)$, then the $\Z_N$ action on $\H$, such that the
quotient $\H/\Z_N$ is the $SU(N)$ singularity,  is given by
\eqn\bino{\left(\matrix{a \cr
b\cr}\right)\to\left(\matrix{e^{2\pi ik/N}a\cr
e^{-2\pi ik/N}b}\right).}
while the $U(1)'$ action that commutes with this (and
preserves the hyper-Kahler structure)
is
\eqn\jpon{\left(\matrix{a \cr
b\cr}\right)\to\left(\matrix{e^{i\psi/N}a\cr
e^{-i\psi/N}b}\right).} The $U(1)'$ action on $\H'$ is similarly
\eqn\hdncon{\left(\matrix{a'\cr
b'\cr}\right)\to\left(\matrix{e^{-i\psi}a'\cr
e^{i\psi}b'}\right).} In all,  if we set $\lambda=e^{i\psi/N}$,
$w_1=\overline a',$  $w_2= b'$, $w_3=a,$ $w_4=\bar b$, then the
quotient $(\H/\Z_N\times \H')/U(1)$ can be described with four
complex variables $w_1,\dots , w_4$ modulo the equivalence
\eqn\nonndon{(w_1,w_2,w_3,w_4)\to
(\lambda^Nw_1,\lambda^Nw_2,\lambda w_3,\lambda
w_4),~~|\lambda|=1.} This quotient is a cone on a weighted
projective space ${\bf WCP}^3_{N,N,1,1}$.  In fact, if we impose
the equivalence relation \nonndon\ for all nonzero complex
$\lambda$, we would get the weighted projective space itself; by
imposing this relation  only for $|\lambda|=1$, we get a cone on
the weighted projective space.

This example was analyzed in \witt. Away from the apex of the
cone, it has a family of $SU(N)$ singularities, giving unbroken
$SU(N)$ gauge symmetry, and an unbroken $U(1)$ carried by the
$C$-field.  The present arguments show that $M$-theory on this
cone has a massless chiral multiplet in the fundamental
representation of $U(N)$, a claim that is entirely compatible
with the anomaly results \witt\ and with the claims (based on a
quite different duality) in section 3.7 of \aw.

Some extensions of this can be worked out in a similar fashion.
Consider the case that away from $P$, the
monodromies break $SU(N+1)$ to $SU(p)\times SU(q)\times U(1)$,
where $p+q=N+1$.  Analysis of the Dirac equation along the above lines
 shows that such a
model will give chiral fermions transforming as $({\bf
p},\bar{\bf q})$ under $SU(p)\times SU(q)$ (and charged under the
$U(1)$).  To describe a dual in  $M$-theory on a manifold of $G_2$
holonomy, we let $K=K'\times U(1)'$, where now $K'=K_1\times K_2$,
$K_1$ being the product of the first $p-1$ $U(1)$'s in $K$ and
$K_2$ the product of the last $q-1$, while $U(1)'$ is the $p^{th}$
$U(1)$. Now we must define $\hat X=\H^{N+1}//K'$, and the manifold
admitting a metric of $G_2$ holonomy should be  $\hat X/U(1)'$.

We can compute $\hat X$ easily, since $K_1$ acts only on
$\Phi_1,\dots,\Phi_p$ and $K_2$ only on
$\Phi_{p+1},\dots,\Phi_{N+1}$.  The hyper-Kahler quotients by
$K_1$ and $K_2$ thus simply construct the $SU(p)$ and $SU(q)$
singularities, and hence $\hat X=\H/\Z_p\times \H/\Z_q$. $\hat X$
has planes of $\Z_p$ and $\Z_q$ singularities, which will persist
in $X=\hat X/U(1)'$, which will also have a more severe
singularity at the origin. So the model describes a theory with
$SU(p)\times SU(q)$ gauge theory and chiral fermions supported at
the origin.
 $U(1)'$ acts on $\H/\Z_p$ and $\H/\Z_q$ as the
familiar global symmetry that preserves the hyper-Kahler
structure of the $SU(p)$ and $SU(q)$ singularities. Representing
those singularities by pairs $\left(\matrix{a \cr b\cr}\right)$
and $\left(\matrix{a'\cr b'\cr}\right)$ modulo the usual action of
$\Z_p$ and $\Z_q$, $U(1)'$ acts by
\eqn\japon{\eqalign{\left(\matrix{a \cr b\cr}\right) &
\to\left(\matrix{e^{i\psi/p}a\cr e^{-i\psi/p}b}\right)\cr
\left(\matrix{a'\cr
b'\cr}\right)&\to\left(\matrix{e^{-i\psi/q}a'\cr
e^{i\psi/q}b'}\right).\cr}} Now if $p$ and $q$ are relatively
prime, we set  $\lambda=e^{i\psi/pq}$, and we find that the
$U(1)'$ action on the complex coordinates $w_1,\dots,w_4$ (which
are defined in terms of $a,b,a',b'$ by the same formulas as
before) is \eqn\mobon{(w_1,w_2,w_3,w_4)\to (\lambda^p
w_1,\lambda^p w_2,\lambda^q w_3,\lambda^q w_4).} If $p$ and $q$
are relatively prime, then the $U(1)'$ action, upon taking
$\lambda$ to be a $p^{th}$ or $q^{th}$ root of 1, generates the
$\Z_p\times \Z_q$ orbifolding that is part of the original
definition of $\hat X$. Hence in forming the quotient $\hat
X/U(1)'$, we need only to act on the $w$'s by the equivalence
relation \mobon.  The quotient is therefore a cone on a weighted
projective space ${\bf WCP}^3_{p,p,q,q}$. If $p$ and $q$ are not
relatively prime, we let $(p,q)=r(n,m)$ where $r$ is the greatest
common divisor and $n$ and $m$ are relatively prime.  Then we let
$\lambda=\exp(i r\psi/pq)$, so \mobon\ is replaced with
\eqn\mobono{(w_1,w_2,w_3,w_4)\to (\lambda^n w_1,\lambda^n
w_2,\lambda^m w_3,\lambda^m w_4).} Also, in addition to imposing
the relation \mobono, to reproduce $\hat X/U(1)$ we must divide
by $\Z_r$, acting by $(w_1,w_2,w_3,w_4)\to (\zeta w_1,\zeta
w_2,w_3,w_4)$, where $\zeta^r=1$.  So $X$ is a cone on ${\bf
WCP}^3_{n,n,m,m}/\Z_r$.  These results were found in section 3.7
of \aw\ using duality with Type IIA brane configurations.

\subsec{ Generalization}

It does not take much more effort to state the generalization.
Suppose that we want to get chiral fermions in the representation
$R$ of a simply-laced group $G$.   This can be achieved for
certain representations.  We find a simply-laced group $\hat G$ of
rank one more than the rank of $G$, such that $\hat G$ contains
$G\times U(1)$ and the Lie algebra of $\hat G$ decomposes as
${\bf g}\oplus {\bf  o} \oplus {\bf r}\oplus \bar {\bf r}$, where
${\bf g}$ and ${\bf o}$ are the Lie algebras of $G$ and $U(1)$,
${\bf r}$ transforms as $R$ under $G$ and of charge 1 under
$U(1)$, and $\bar{\bf r}$ transforms as the complex conjugate.
Such a  $\hat G$ exists only for special  $R$'s, and these are
the $R$'s that we will generate from $G_2$ singularities.

Given $\hat G$, we proceed as  above on the heterotic string side.
We consider a family of $\T^3$'s, parameterized by $Q$, with
monodromy that at a special point $P\in Q$ leaves unbroken $\hat
G$, and at a generic point breaks $\hat G$ to $G\times U(1)$.  We
moreover assume that near $P$, the monodromies have the same sort
of generic behavior assumed above.  Then the same computation as
above will show that the heterotic string has, in this situation,
a single multiplet of fermion zero modes (of positive or negative
chirality depending on the sign of the analog of
$\sigma_1\sigma_2\sigma_3$) in the representation $R$, with
$U(1)$ charge 1.

Dualizing this to an $M$-theory description, over $P$ we want a
$\hat G$ singularity, while over a generic point in $Q$, we
should have a $G$ singularity.  Thus, we want to consider the
unfolding of the $\hat G$ singularity (as a hyper-Kahler
manifold) that preserves a $G$ singularity.  To do this is quite
simple.  We start with the Dynkin diagram of $\hat G$.  The
vertices are labeled with integers $n_i$, the Dynkin indices.  In
Kronheimer's construction, the $\hat G$ singularity is obtained
as the hyper-Kahler quotient of $\H^k$ (for some $k$) by the
action of a group $K=\prod_i U(n_i)$. Its unfolding is obtained
by allowing the $\vec D$-fields of the $U(1)$ factors (the
centers of the $U(n_i)$) to vary.

The $G$ Dynkin diagram is obtained from that of $\hat G$ by
omitting one node, corresponding to one of the $U(n_i)$ groups;
we write the center of this group as $U(1)'$.  Then we write $K$
(locally) as $K=K'\times U(1)'$, where $K'$ is defined by
replacing the relevant $U(n_i)$ by  $SU(n_i)$. We get a
hyper-Kahler eight-manifold as the hyper-Kahler quotient $\hat
X=\H^k//K'$, and then we get a seven-manifold $X$ by taking the
{\it ordinary} quotient $X=\hat X/U(1)'$.  This maps to $Q=\R^3$
by taking the value of the $U(1)'$ $\vec D$-field, which was not
set to zero. The fiber over the origin is obtained by setting
this $\vec D$-field to zero after all, and gives the original
$\hat G$ singularity, while the generic fiber has a singularity
of type $G$.

Just as in \kv, one can readily work out examples of pairs $G, $
$\hat G$. We will just consider the cases most relevant for grand
unification. For $G=SU(N)$, to get chiral fields in the
antisymmetric tensor representation, $\hat G$ should be $SO(2N)$.
For $G=SO(10)$, to get chiral fields in the ${\bf 16}$, $\hat G$
should be $E_6$. For $G=SO(2k)$, to get chiral fields in the
${\bf 2k}$,  $\hat G$ should be $SO(2k+2)$. (Note in this case
that ${\bf 2k}$ is a real representation. However, the
monodromies in the above construction break $SO(2k+2)$ to
$SO(2k)\times U(1)$, and the massless ${\bf 2k}$'s obtained from
the construction are charged under the $U(1)$;  under
$SO(2k)\times U(1)$ the representation is complex.) For $2k=10$,
this example might be used in constructing $SO(10)$ GUT's.   For
$G=E_6$, to get ${\bf 27}$'s, $\hat G$ should be $E_7$.  A useful
way to describe the topology of $X$ in these examples is not
clear.

In this construction, we emphasized, on the heterotic string
side, the most generic special monodromies that give enhanced
gauge symmetry, which corresponds on the $M$-theory side to
omitting from the hyper-Kahler quotient a rather special $U(1)$
that is related to a single node of the Dynkin diagram.  We could
also consider more general heterotic string monodromies; this
would correspond in $M$-theory to omitting a more general linear
combination of the $U(1)$'s.

\subsec{Generation Of A Superpotential}

Finally, let us briefly indicate how to map the superpotential as computed
on the heterotic string side to an $M$-theory computation.

In heterotic string compactification on a Calabi-Yau threefold $W$,
the chiral fermion wavefunctions are spread out over $W$.  The superpotential
has a classical contribution, coming from the integral of a product of chiral
wavefunctions, and corrections from worldsheet instantons of genus zero.

In the present example, $W$ has a map to a three-manifold $Q$.
Worldsheet instantons cannot project to a point in $Q$ (the fiber
is Lagrangian and does not contain any holomorphic curves), so
they must project to one-surfaces or two-surfaces in $Q$. A
worldsheet instanton that projects to a two-surface in $Q$ would
come in $M$-theory from a fivebrane instanton (since macroscopic
heterotic strings in the seven-dimensional theory obtained by
compactification on $\T^3$ are dual to strings made from
fivebranes wrapped on K3).  There are no supersymmetric
fivebranes on a $G_2$-manifold $X$ (in fact $b_1(X)=0$) so we
conclude, as could presumably be argued directly, that the
relevant heterotic string instantons project to one-manifolds
(possibly singular one-manifolds, i.e. graphs) in $Q$.

A worldsheet instanton that projects to a one-manifold describes the propagation
of a string wrapped on a one-cycle in the $\T^3$ fiber.  Such a string lifts
in $M$-theory to a membrane wrapped on a two-cycle in K3.  So worldsheet
instantons in the heterotic string correspond to membrane instantons in $M$-theory.

What about the classical contribution from the heterotic string?
At first sight, one might expect that this would come from a
classical contribution on the $M$-theory side.  But, with each
$G$-multiplet of chiral fields supported at a distinct
singularity, it is very hard to construct anything in $M$-theory
that one might regard as a ``classical'' contribution.  (There is
no superpotential term that can be constructed locally involving
only chiral fields supported at just one singularity, because
they are always in an irreducible representation of the gauge
group with a fixed and nonzero charge under a $U(1)$ that is
unbroken locally.)
 We will argue, instead, that also the classical contribution
on the heterotic string side corresponds in $M$-theory to a
membrane instanton sum.

The idea is that on the heterotic string side, the chiral
multiplets $\Phi_\alpha$
 are supported at points $P_\alpha\in Q$ that we assume to be disjoint.
The wavefunction for each $\Phi_\alpha$ decays exponentially in
the distance from $P_\alpha$, so in the limit that the fibers of
$W\to Q$ have small volume, the classical contributions are
exponentially suppressed (and hence can come from membrane
instantons on the $M$-theory side). To compute a classical
contribution to a term in the superpotential of the form, say,
$\Phi_\alpha\Phi_\beta\Phi_\gamma$ for distinct
$\alpha,\beta,\gamma$, one would find a graph $\Gamma$ in $Q$ of
``steepest descent'' connecting $P_\alpha$, $P_\beta$, and
$P_\gamma$, in which the exponential decay of the wavefunctions
is minimized.  Each line in the graph represents propagation of a
mode of one of the $\Phi$'s.  The components of $\Phi$ are
charged strings; they can be regarded as strings wrapped, not on
the geometrical fiber $\T^3$ of $W$, but on the ``internal
torus'' $\T^{16}=\T^8\oplus \T^8$ that carries the charged
degrees of freedom in the bosonic construction of the heterotic
string. Such internally wrapped strings also correspond to
wrapped membranes in $M$-theory, so their propagation is again
described in $M$-theory by a membrane instanton effect.

One detail should be checked here, namely that the wrapped
membranes that correspond to the $\Phi$'s are not collapsed at
singularities.  The relevant chiral modes of the $\Phi$'s are in
the adjoint representation of $\widehat G$, and not in the
adjoint representation of $G$.  Once one leaves the points
$P_\alpha$, the $M$-theory singularities are of type $G$, and only
membranes that represent charges in the adjoint representation of
$G$ are collapsed at singularities.

The fact that the propagating strings are charged under $G$ means
that the corresponding wrapped membranes pass through the orbifold
singularity.  Hence, the intersection of the membrane instanton
with $Q$ is a graph connecting the points $P_\alpha$, just as we
observed in the heterotic string description.

One could consider a more special situation in which on the
heterotic string side several of the $P_\alpha$ coincide.  Then
in $M$-theory there would be some more complicated singularities
giving several  gauge multiplets of chiral fields; in this
situation some of the membrane instantons might collapse and some
terms in the superpotential might not have a membrane instanton
interpretation.  An example is the cone on $SU(3)/U(1)^2$
considered in \aw; it gives three gauge multiplets of chiral
fermions, with a superpotential that presumably cannot usefully
be derived from membrane instantons.

\newsec{Singularities from Twistor Spaces}

The conical $G_2$ manifolds whose existence was argued in the
previous section were of the form $\hat X/U(1)$, where $\hat X$ is
a conical hyperkahler eight-manifold that is obtained from a
hyper-Kahler quotient, and $U(1)$ preserves the hyper-Kahler
structure of $\hat X$. These examples had the advantage that we
know (via duality with the heterotic string) exactly what the
physics is, but the disadvantage that we do not know how to
actually construct a $G_2$ metric on $\hat X/U(1)$.

Here we will consider a different quotient of the form $\hat
X/U(1)$, with the same $\hat X$ but a different $U(1)$ action. The
properties will be opposite: the physics is in general harder to
understand, but a conical $G_2$-metric on $\hat X/U(1)$ can be
constructed explicitly. In this case, the $U(1)$ action on $\hat
X$ does {\it not} preserve the hyper-Kahler structure of $\hat X$;
it is part of an $SU(2)$ that rotates the three complex
structures.

Let us first recall why $\hat X$ has such an $SU(2)$ action. (It
also entered briefly in section 2.)  Existence of this $SU(2)$ is
inherited from the fact that $\hat X$ is a hyper-Kahler quotient
of a system of $n$ free hypermultiplets, for some $n$. The bosons
in these hypermultiplets parameterize ${\bf R}^{4n}$ for some
$n$. ${\bf R}^{4n}$ admits the action of $SU(2)\times Sp(n)$,
where $SU(2)$ rotates the three complex structures and $Sp(n)$
preserves them. To obtain $\hat X$, we pick a subgroup $H$ of
$Sp(n)$ (the dimension of $H$ being $n-2$), and form the
hyper-Kahler quotient -- that is, we divide by $H$ and set to
zero the $D$-fields (or hyper-Kahler moment map) of $H$.  These
operations are $SU(2)$-invariant, so the hyper-Kahler quotient
$\hat X={\bf R}^{4n}//H$ inherits an $SU(2)$ action that rotates
the three complex structures.

Now we simply pick a $U(1)$ subgroup of $SU(2)$, and we claim
that the quotient $X=\hat X/U(1)$ has a conical $G_2$ metric which
moreover can be described explicitly.

To show this requires several steps.    $\hat X$ is a cone on a
seven-manifold $V$, and $V$ also admits the action of $SU(2)$.
($V$ has a tri-Sasakian structure \threesas\
 with $SU(2)$ rotating the
three vector fields.) $SU(2)$ acts on $V$ with three-dimensional
orbits, so the quotient $V/SU(2)$ is, away from some possible
singularities, a four-manifold $M$. The metric that $M$ inherits
as the quotient of $V$ is actually a self-dual Einstein metric of
positive scalar curvature.  (The only smooth manifolds of this
type are $\S^4$ and $\CP^2$, so $M$ is almost always singular;
the singularities arise because $SU(2)$ does not act freely on
$V$. It can be shown that the singularities of $M$ are orbifold
singularities.)

One way to explain that the metric on $M$ is a self-dual Einstein
metric is as follows. $\R^{4n}$ to begin with is a cone on a
sphere $\S^{4n-1}$, and the quotient $\S^{4n-1}/SU(2)$ is a copy
of quaternionic projective space $\HP^{n-1}$, with a metric of
$Sp(1)\cdot Sp(n-1)$ holonomy.  For manifolds of such holonomy
there is a quaternionic-Kahler quotient construction
\ref\galaw{K. Galicki and H. Lawson,``Quaternionic Reduction and
Quaternionic Orbifolds'' Math. Ann. {\bf 282}  (1988) 1}
analogous to the hyper-Kahler quotient.  $M$ can be characterized
as the quaternionic-Kahler quotient of $\HP^{n-1}$ by $H$, and
hence inherits a quaternionic-Kahler structure, which for a
four-manifold is the same as a self-dual Einstein metric of
positive scalar curvature.

Now let us describe $\hat X/U(1)$.  For any conical manifold such
as $\hat X$ or $\R^4$, we let $\hat X_0$ or $\R^4_0$ denote the
same space with the origin deleted.  For example, $\hat
X_0=\R^+\times V$, where $ \R^+$ is the positive real numbers.
$V$ is an $SU(2)$ bundle over $M$, so $\hat X_0$ is an
$\R^+\times SU(2)$ bundle over $M$, or equivalently an $\R^4_0$
bundle over $M$ (since $\R^4_0=\R^+\times SU(2)$).  $SU(2)$ acts
on the $\R^4_0$ fibers via its standard action on $\R^4$, and the
$U(1)$ we want to divide by is a subgroup of this $SU(2)$. With
this information it is easy to describe $X_0=\hat X_0/U(1)$. Since
$\R^4/U(1)=\R^3$ (this being the basis for the standard lift of
the $D6$-brane to $M$-theory \ref\townsend{P. Townsend, ``The
eleven-dimensional supermembrane Revisited,'' Phys. Lett. {\bf
B350} (1995) 184. hepth/9501068.}), we have $\R^4_0/U(1)=\R^3_0$,
and hence $X_0 $ is an $\R^3_0$ bundle over $M$.

Next we want to include the origin in $\hat X$, and there are two
natural-looking ways to do this.  The origin is a single point in
$\hat X$, so it descends to a single point in $\hat X/U(1)$.  So,
to describe $\hat X/U(1)$, we should add just one point to
$X_0$.  To do so, we note that $\R^3_0=\R^+\times \S^2$, so an
$\R^3_0$ bundle over $M$, such as $X_0$, is asymptotic at
infinity to a cone on an $\S^2$ bundle over $M$; we will let $Y$
denote this $\S^2$ bundle. Then, if we let $X$ be a cone on $Y$,
$X$ differs from
 $X_0$ precisely by adding a single point at the origin, and so
$X=\hat X/U(1)$.

Alternatively, we could compactify the interior of $\hat X_0$ by
replacing the $\R^3_0$ bundle over $M$ by an $\R^3$ bundle over
$M$, simply by filling in the ``origin'' in each fiber.  This
gives another seven-manifold, which we will call $X'$.

In fact, it can be shown that $X'$ is the bundle of
anti-self-dual two-forms over $M$, sometimes denoted
$\wedge^{2,-}(M)$.  $Y$ is the bundle of unit anti-self-dual
two-forms, also known as the twistor space of $M$.

Since $M$ is a self-dual Einstein manifold,  a general
construction \refs{\brysal - \gibb} gives $G_2$ holonomy metrics
on the bundles $X$ and $X'$ over $M$. The $G_2$ metric on $X'$
can be written explicitly in terms of the metric $d{\Sigma}^2$ on
$M$, fiber coordinates $u_i$ for the bundle $X' \to M$, together
with the Riemannian connection $A$ on the bundle of
anti-self-dual two-forms: \eqn\nonp{ds^2={dr^2\over 1-\left({r_0/
r}\right)^4}+{r^2\over 4} (1-(r_0/r)^4)|d_Au|^2+{r^2\over
2}d{\Sigma}^2} Here $d_Au$ is the covariant derivative
$d_Au_i=du_i+\epsilon_{ijk} A_j{u_k}$. $r_0$ is the modulus of
the solution; it determines the volume of the copy of $M$ at the
center of $X'$.  As $r_0\to 0$, $X'$ degenerates to $X$, and
\nonp\ degenerates to a conical $G_2$ metric on $X$.

The asymptotically conical metric \nonp\ has also been described
in \ref\reyes{R. Reyes-Carion, ``Special Geometries Defined By Lie
Groups,'' Ph.D thesis, Oxford, 1993.} starting from the canonical
metric on the twistor space.

\subsec{Determination Of $M$ In Some Examples}

This construction can be used to construct many possible examples.
To make it concrete, one would like an effective way to describe
$M$.  This has been done in \twist\ for a simple class of
examples; the classical singularities of these examples were also
investigated in \twist.  In the rest of this paper, we will
explain these results and then attempt to analyze the behavior of
these examples in $M$-theory.

For these examples, $H = U(1)$ and there are three
hypermultiplets $\Phi_i$ of charges  ${p_1},{p_2},{p_3}$.   We
write the $\Phi_i$ in terms of complex fields $a_i,\overline
b_i$. By possibly exchanging some $a$'s and $b$'s, we can assume
that the $p_i$ are all positive. Also, by possibly rescaling the
generator of $U(1)$, we can assume that the $p_i$ have no common
factor.

The hyper-Kahler moment map can be written in terms of a real part
\eqn\rmom{{\mu}_{\R} (p_i ) = \sum_i p_i ( |a_i|^2 - |b_i|^2 )}
and a complex part \eqn\cmom{{\mu}_{\C} (p_i) = \sum_i p_i  a_i
{\bar b}_i .} If we introduce $x_i=\sqrt{p_i}a_i$ and $y_i={\sqrt
p_i}b_i$, then in terms of the new variables \rmom\ and \cmom\
become \eqn\rmomb{{\mu}_{\R}({\bf 1}) = \sum_i ( |x_i|^2 -
|{y_i}|^2 )} and \eqn\cmomb{{\mu}_{\C} ({\bf 1)} = \sum_i x_i
{\bar y}_i .} This rescaling commutes with both $SU(2)$ and $H$.
The submanifold $N$ of ${\H}^3$ on which the hyper-Kahler moment
map vanishes coincides with the zero set of \rmomb\ and \cmomb\ .
$N$ is in fact a cone on $SU(3)$.

To see this, think of an element $U\in SU(3)$ as a $3\times 3$
complex matrix whose columns are elements ${\bf u},$ ${\bf v}$,
and ${\bf w}$  of $\C^3$.  The fact that $U$ is unitary means
that ${\bf u}$, ${\bf v}$, and ${\bf w}$ are orthonormal.  The
fact that $U$ has determinant one means that ${\bf u}\times {\bf
v}\cdot {\bf w}=1$, where here $\times$ is the usual cross product
of three-vectors, and $\,\,\cdot\,\,$ is the usual dot product of
three-vectors.  Given these conditions, ${\bf w}$ is uniquely
determined in terms of ${\bf u}$ and ${\bf v}$; in fact, ${\bf
w}=\overline {\bf u}\times \overline{\bf v}.$ Hence elements of
$SU(3)$ are in one-to-one correspondence with pairs ${\bf u},{\bf
v}\in \C^3$ with $|{\bf u}|^2=|{\bf v}|^2=1$, ${\bf u}\cdot
\overline{\bf v}=0$. Modulo scaling of ${\bf x}$ so that $|{\bf
x}|^2=1$, these are precisely the conditions \rmomb\ and \cmomb\
obeyed by ${\bf x}$ and ${\bf y}$.  So the space of zeroes of
\rmomb\ and \cmomb\ is a cone on $SU(3)$.

The manifold $SU(3)$ also admits an action of $SU(2)$ rotating the
first two columns ${\bf u}$ and ${\bf v}$ among themselves and
leaving the third column ${\bf w}$ fixed. This is the $SU(2)$
that rotates the complex structures of $\hat X$.  It is easy to
take the quotient $SU(3)/SU(2)$ in this language.  Given three
orthonormal vectors ${\bf u}$, ${\bf v}$, ${\bf w}$ in $\C^3$ with
${\bf u}\cdot {\bf v}\times {\bf w}=1$,  the choice of ${\bf w}$
uniquely determines ${\bf u}$ and ${\bf v}$ up to an $SU(2) $
transformation.  So we can divide by $SU(2)$ by just forgetting
${\bf u}$ and ${\bf v}$.  So the quotient $SU(3)/SU(2)$ is a copy
of $\S^5$, parameterized by a complex three-vector ${\bf w}$ with
$|{\bf w}|^2=1$.  Such a ${\bf w}$ determines an element of
$\S^5$; $N/SU(2)$ is a cone on this $\S^5$.

 To get the
 desired four-manifold $M$, we must divide this ${\bf S}^5$ by the
 original $U(1)$ that acts on the hypermultiplets with charges
 $p_1,p_2,p_3$.
Since $H$ acts on the components of $\bf x$ and $\bf y$  with
eigenvalues $({p_1},{p_2},{p_3})$,  it acts on the components of
$\bf w=\overline {\bf x}\times \overline {\bf y}$ with eigenvalues
$-({p_2 + p_3 , p_1 + p_3 , p_1 + p_2 })$. The overall minus sign
can be removed by reversing the sign of the generator of $H$.
After doing this, the $U(1)$ action on the $w_i$ is
 \eqn\toon{w_i \to
{\lambda}^{k_i} w_i} where $(k_1 , k_2 , k_3 )$ $=$ $(p_2 + p_3 ,
p_1 + p_3 , p_1 + p_2 )$. Since we assumed that the $p_i$ have no
common factor, the $k_i$ either have no common factor or have a
greatest common divisor of 2; the latter possibility arises
precisely if the $p_i$ are all odd.

$M$ is the quotient ${\S}^5 /H$ and is therefore a copy of
${\WCP}^2_{q_1,q_2,q_3} $, where the weights $q_i$ are $k_i$ if
the $p_i$ are not all odd, and are $k_i/2$ if the $p_i$ are all
odd.

\subsec{The Singularities}

We want to understand the physics of $M$ theory on the singular
$G_2$-holonomy spaces derived from the examples just considered.
We must first understand the singularities of these spaces. We
begin by describing the singularities of $M={\bf
WCP}^2_{{q_1},{q_2},{q_3}}$.  We continue to denote the
homogeneous coordinates of this weighted projective space as
$w_i$.

Although the $q_i$ have no common factor, they are not necessarily
{\it pairwise} relatively prime. Let $n_1$ be the greatest common
divisor of $q_2$ and $q_3$, and similarly for $n_2$ and $n_3$. If
we set $w_i=0$, then $\lambda$ acts trivially in \toon\ if
$\lambda^{n_i}=1$, so we get a $\Z_{n_i}$ orbifold singularity.
Setting $w_i=0$ leaves us with a copy of $\CP^1$ (note that a
weighted $\CP^1$ is equivalent topologically to an ordinary one).
So in general, if the $n_i$ are all greater than one, we have
three $\CP^1$'s of orbifold singularities in $M$. The $i^{th}$ and
$j^{th}$ such $\CP^1$'s meet at the point with $w_i=w_j=0$. The
configuration is thus a ``triangle'' of $\CP^1$'s.  The vertices
of the triangle -- where $w_i$ and $w_j$ vanish for some $i$ and
$j$ -- are orbifold singularities of order $q_k$ where $k$ is the
``third'' label.

Now let us analyze the singularities in the bundle $X'$ of
anti-self-dual two-forms over $M$ and in the twistor space $Y$ of
unit anti-self-dual two-forms.  To find a fixed point in $X'$ or
$Y$, we have to find a point in $M$ that is invariant under
\eqn\kili{w_i\to \lambda^{q_i}w_i} (for some value of
$\lambda\not= 1$) and an anti-self-dual two-form that is also
invariant. So the key issue is to determine how \kili\ acts on the
anti-self-dual two-forms over the orbifold singularities in $M$.

Suppose, for example, we consider the fixed $\CP^1$ with $w_1=0$.
Near this locus the anti-self-dual two-forms can be written
explicitly \eqn\kopo{dw_1\wedge d\bar w_1 -dt\wedge d\bar t,~~
dw_1\wedge d\bar t,~~ dt\wedge d\bar w_1,} where $t$ is a linear
combination of $w_2$ and $w_3$.  For $\lambda^{n_1}=1$, these
anti-self-dual two-forms transform by
$1,\lambda^{q_1},\lambda^{-q_1}$, respectively.  Since
$\lambda^{q_1}\not=1$ if $\lambda\not= 1$ (given that the $q_i$
have no common factor) these eigenvalues are those of a
nontrivial rotation of $\R^3$.  The fixed point set in $\R^3$
consists of a single line -- the multiples of $\theta=dw_1\wedge
d\bar w_1-dt\wedge d\bar t$.  In $\S^2$, the fixed point set
consists of two points -- the points where the fixed line in
$\R^3$ meets $\S^2$.

So in $X'$, the singular set consists of three copies of
$\CP^1\times \R$, the $i^{th}$ copy being an orbifold singularity
of order $n_i$.  They meet pairwise at a copy of $\R$ at which two
of the $w_i$ vanish.  On these sets, we again get an orbifold
singularity, of higher order.  For example, consider the locus
with $w_1=w_2=0$, $w_3=1$, so to get a fixed point we need
$\lambda^{q_3}=1$. The anti-self-dual two forms in this locus are
$|dw_1|^2-|dw_2|^2$, $dw_1\wedge d\bar w_2$, $d\bar w_1\wedge
dw_2$ and transform as $1$ and $\lambda^{\pm (q_1-q_2)}$.  The
number of values of $\lambda$ for which
$\lambda^{q_3}=\lambda^{q_1-q_2}=1$ is $r_3$, the greatest common
divisor of $q_3$ and $q_1-q_2$.  So if $r_3>1$,  the point with
$w_1=w_2=0$ lifts in $X'$ to an $\R^3$ of $\Z_{r_3}$ orbifold
singularities.  If $r_3=1$, then to get a fixed point, we must
take the anti-self-dual two-form to be a multiple of
$|dw_1|^2-|dw_2|^2$, so in this case the point with $w_1=w_2=0$
lifts to a fixed $\R$ in $X'$.  This copy of $\R$ is a locus of
$\Z_{q_3}$ orbifold singularities; the action of $\Z_{q_3}$ on
the normal bundle can be described by saying that the normal
bundle is a copy of $\C^3$ and the eigenvalues are
$\lambda^{-q_1}, \lambda^{q_2},$ and $\lambda^{q_1-q_2}$ (acting
on $\bar w_1$, $w_2$, and an anti-self-dual two-form). In the
twistor space $Y$, the story is similar, except that we want unit
anti-self-dual two-forms, so a fixed $\R^3$ is replaced by $\S^2$
and a fixed $\R$ by two points.  Of course, there is a similar
story to the above with $q_3$ replaced by any of the $q_i$ and
$r_i$ defined as the greatest common divisor of $q_i$ and
$q_{i+1}-q_{i-1}$.

For $\Z_n$ fixed points of codimension four (of the form
$\CP^1\times\R$ or $\R^3$), the action on the normal bundle is
always that of a standard $A_{n-1}$ singularity (the eigenvalues
being $q^{\pm 1}$ where $q^n=1$).  This is required for $G_2$
holonomy (or supersymmetry) and can be verified from the above
formulas.

As $X$ is a cone on $Y$, the fixed points in $X$ are just a cone
on the fixed point set in $Y$, so each fixed $\S^2$ (or fixed
point) is replaced by a fixed $\R^+\times \S^2$ (or $\R^+$). Of
course $\R^+$ is topologically the same as $\R$, but it is
natural to here think of a half-line.  What happens to the fixed
point set when the conical manifold $X$ is deformed to the
orbifold $X'$, the bundle of anti-self-dual two-forms?  The two
$\R^+\times \S^2$'s over a fixed $\CP^1$ in $M$ join together in
$X'$ to a fixed $\R\times \S^2$, by gluing two half-lines $\R^+$
into a copy of $\R$. As for fixed $\R^+\times \S^2$'s that lie
over vertices of the triangle (when some $r_i$ are greater than
one), here we note that upon adding an ``origin,'' $\R^+\times
\S^2$ is the same as $\R^3$; these give the fixed $\R^3$'s in
$X'$ that were found above.

\subsec{$M$-Theory Physics on $X$ and $X'$}

We are now in a position to say something about the $M$-theory
physics near these $G_2$-holonomy singularities. We begin by
discussing the abelian symmetries associated with the $C$-field.
We then explain why there must be chiral fermions charged under
these symmetries supported at the conical singularity in the cone
on $Y$. Following this we go on to discuss the physics associated
with the additional orbifold singularities.

\bigskip
\noindent{\it $C$-field symmetries and Chiral Fermions.}

In $M$-theory on $X$ with $X$ a manifold with $G_2$-holonomy, the
Lie algebra of the symmetries which originate from the $C$-field
is isomorphic to ${H^2}(X; {\R})$, the second de Rham cohomology
group. If $Y$ is the twistor space of   ${\bf
WCP}^2_{{q_1},{q_2},{q_3}}$, then the second Betti number of $Y$
is two. This is also the second Betti number of the cone on $Y$.
Thus in $M$-theory on the cone on $Y$ we have, at least locally, a
$U(1)^2$ symmetry group originating from the $C$-field.

On the other hand, the smoothed out cone $X'$ is contractible to
${\bf WCP}^2_{{q_1},{q_2},{q_3}}$, and therefore its second Betti
number is one. Thus, when we deform the cone, the $U(1)^2$
symmetry gets broken to $U(1)$. This is exactly as in the case
studied in \aw\ when all the weights are one. As will become
apparent below, other symmetries of the physics also get broken
when we deform the cone.

In \witt, it was shown that in $M$-theory on the cone on $Y$ we
will find chiral fermions charged under $U(1)^2$ if \eqn\anom{
{\int}_{Y} {\omega}_i \wedge {\omega}_j \wedge {\omega}_k \neq 0
} for any $\omega_i\in{H^2}(Y;{\Z})$. By taking two $\omega$'s to
be pullbacks from the base and the third to have a nonzero
integral over the fiber, we can make \anom\ nonzero. This works
for any $M$ of positive second Betti number (such as the examples
considered here) and implies that  there are chiral fermions,
charged under the abelian gauge symmetries from the $C$-field,
and supported at the conical singularity.

We now go on to consider the physics associated with the
singularities in the cone on $Y$ and its smoothed out version.
The general discussion is rather complicated, as one might guess
from the discussion of the singularities above, and we will focus
on certain classes of examples.  For simplicity, we will mainly
consider cases in which the $r_i$ are all 1.

\bigskip \noindent{\it The Generic Case}

We will first consider the case, which one might loosely consider
generic, in which the $q_i$ are pairwise relatively prime as well
as $r_i=1$.  In this case, $M$ has three isolated orbifold points
(where two of the $w_i$ vanish), and $Y$ contains six isolated
orbifold points. In the neighbourhood of the $i^{th}$ such point,
$Y$ is locally ${{\C}^3}/{\Z_{q_i}}$ and as found above,
$\Z_{q_i}$ acts with eigenvalues
$\lambda^{-q_{i+1}},\lambda^{q_{i-1}},\lambda^{q_{i+1}-q_{i-1}}$.

There is not a known useful description of the behavior of
$M$-theory at a codimension six orbifold singularity of this type.
 It is believed that   $M$-theory on
$({\C^3}/{\Z_{q_i}})\times \R^5$ with generic action of
$\Z_{q_i}$ generates at low energies a non-trivial conformal
theory in five dimensions.  (This assertion generalizes special
cases that have been studied in \nref\dkv{M. R. Douglas and S.
Katz, and C. Vafa, ``Small Instantons, del Pezzo Surfaces, and
Type
I$'$ Theory, hep-th/9609071.}%
\nref\ms{D. R. Morrison and N. Seiberg, ``Extremal Transitions
And Five-Dimensional Supersymmetric Field Theories,'' Nucl. Phys.
{\bf B483} (1997) 229, hep-th/9609070; K. Intriligator, D. R.
Morrison, N. Seiberg, ``Five-Dimensional Supersymmetric Gauge
Theories And Degenerations Of Calabi-Yau Spaces,'' Nucl. Phys.
{\bf 497B} (1997) 56, hep-th/9702198.}%
\nref\ewt{E. Witten, ``Phase Transitions in $M$ Theory and $F$
Theory,'' Nucl. Phys. {\bf B471} (1996) 195, hep-th/9603150.}%
\refs{\dkv - \ewt}.)

Assuming that this is correct, the low energy behavior of
$M$-theory on $\R^4\times X'$ is  determined by those conformal
field theories.  The three fixed lines in $X'$ correspond to
copies of $\R^4\times \R=\R^5$ on which nontrivial conformal
field theories are realized.  There are no further complications
because the fixed $\R$'s in $X'$ do not intersect.

If we degenerate the orbifold $X'$ to the cone $X$, things are
more complicated as the $\R$'s intersect at the conical
singularity at the origin.  Additional phenomena might be
expected at this point, especially since a chiral anomaly is
supported there, as we have seen above.

\bigskip
\noindent{\it A Special Case}

Having discussed a general case where we can say very little, we
will now go on to discuss a case which we can more or less
understand fully. This is when the $q_i$ take the form $(p,p,q)$
with $p$ and $q$ relatively prime. In this case, $\Sigma$, the
singular set of $Y$, consists of a pair of two-spheres of
$A_{p-1}$ singularities and a single two-sphere of
$A_{q-1}$ singularities. In the cone on $Y$ each of these becomes
a copy of ${\R}^3$ and all three  meet at the origin.  The three
$\R^3$'s support gauge groups that are respectively $SU(p)$,
$SU(p)$, and $SU(q)$.  Locally the topology of this singularity is
very similar to a case discussed in \aw\ in which three ${\R}^3$
families of $A_{n-1}$ singularities met at the conical
singularity. In addition to the $SU(n)^3$ gauge symmetry it was
found that chiral fermions were residing at the origin in the
representation ${\bf (n,{\bar n},1)} + {\bf (1,n,{\bar n})} +
{\bf ({\bar n},1,n)}$. In the example we are discussing here, the
only difference is that one of the ${\R}^3$'s of singularities is
of a different rank from the other two. We therefore propose that
in this example there are chiral fermions at the origin in the
representations ${\bf (p,{\bar p},1)} + {\bf (1,p,{\bar q})} +
{\bf ({\bar p},1,q)}$.

This proposal is compatible with the anomaly calculations in
\witt. We can argue for it more precisely by relating the problem
to a Type IIA configuration on $\R^6$, with  three sets of
intersecting D6-branes (each filling out a copy of $\R^3$ passing
through the origin) of multiplicities $(p,p,q)$.

 With ${\bf WCP}^2_{p,p,q}$ defined via the $U(1)$  action on ${\C}^3$
 that multiplies the homogeneous coordinates by
 $({{\lambda}^p},{{\lambda}^p},{{\lambda}^q})$, consider the $U(1)$ action
 $\alpha$ that acts by  by $({\tau},{\tau},1)$, $|\tau|=1$.
 The fixed points of $\alpha$ in the cone on $Y$ are precisely
 the orbifold singularities $\Sigma$. Proceeding
 exactly as in section $3.4$ of \aw\
 one can show that $Y/{U(1)}={\S}^5$ and that the corresponding Type IIA description is in
 terms of D6-branes in flat space. \foot{For more general
 weights the reasoning given in \aw\ does not hold and it is unlikely
 that one can relate the corresponding $M$-theory physics to $D$-branes in flat space.}

 The description in terms of branes makes it
 apparent that there is a cubic superpotential between the three chiral multiplets. Consider now
 Higgsing the $SU(p)^2$ gauge symmetry down to its diagonal $SU(p)$ by giving a vev
 to the $({\bf p, {\bar p}, 1})$
 superfield. The superpotential implies that the remaining two fields become massive so that the
 only remaining
 massless degrees of freedom are the $SU(p)$ and $SU(q)$ vector multiplets.\foot{Here we have used
 a terminology that is strictly appropriate if the singularity is embedded in a compact manifold
 and four-dimensional effective field theory can be used.} This is exactly in accord
 with our expectation of the physics in the smoothed out cone. There the singularities consist of an
 ${\R}{\times}{\S}^2$ of
 $A_{p-1}$-singularities and an ${\R}^3$ of $A_{q-1}$-singularities which do not intersect.
 Therefore, again we see that
 deforming the cone corresponds to symmetry breaking.

\bigskip
\noindent{\it  Massless Hypermultiplets}

Now we want to consider the situation in which the $q_i$ are not
relatively prime.  In general, one can meet a mixture of all the
phenomena explained above and below, but if the $q_i$ are of the
form $(ab,ac,bc)$ for some relatively prime integers $a,b,c$,
there is an additional phenomenon that can be described in a
simple fashion. So we will focus on this case.

With our choice, the greatest common divisors of pairs of $q_i$
are $a,$ $b$, and $c$, respectively.  So the fixed $\CP^1$'s in
${\bf WCP}^2_{ab,ac,bc}$ are $\Z_a$, $\Z_b$, and $\Z_c$
singularities while their intersections are $\Z_{ab}$, $\Z_{bc}$,
and $\Z_{ca}$ singularities. In $Y$, everything is duplicated. In
a neighbourhood of one of the orbifold points in $Y$ labeled by
$ab$, the orbifold singularity looks like the origin in
${{\C}^3}/{{\Z_a}{\times}{\Z_b}}$. Specifically, ${\Z_a}$ acts on
local coordinates of $\C^3$ by multiplication by $(q, q^{-1},1)$,
$q^a=1$, and $\Z_b$ acts by $(1,q,q^{-1})$, $q^b=1$.  This is a
special case of what was explained in section $3.2$.

\nref\vbs{M. Bershadsky, C. Vafa, and V. Sadov, ``D-strings on
D-manifolds,'' Nucl. Phys. {\bf B463} (1996) 398.
hepth/9510225.}%
The physics of $M$-theory on $\R^5\times
({{\C}^3}/{{\Z_a}{\times}{\Z_b}})$ is believed to be described by
a five dimensional supersymmetric gauge theory with
$SU(a){\times}SU(b)$ gauge group and a single bi-fundamental
hypermultiplet \refs{\vbs,\kv} that is supported at the origin.
Therefore, in $M$-theory on  $\R^4\times \wedge^{2,-}({\bf
WCP}^2_{ab,ac,bc})$, along with the gauge groups supported on
copies of $\R^4\times \S^2\times \R$, there are bi-fundamental
hypermultiplets supported on the intersections $\R^4\times \R$,
transforming as  ${(\bf a,{\bar b},1) + (1,b,{\bar c}) + ({\bar
a}, 1, c)}$ $+c.c$.

A cone $X$ on $Y$ has more complicated singularities, so we do
not immediately learn how to describe the spectrum in the case of
the cone. If we think of $X'=\wedge^{2,-}({\bf WCP}^2_{ab,ac,bc})$
as a deformation of the cone, can we guess the physics on $X$ by
requiring that it reduce to the physics on $X'$ after symmetry
breaking?   We will not be able to offer a satisfactory answer to
this question but we will offer a suggestion.

When $X'$ degenerates to $X$, each fixed $\CP^1\times \R$ splits
into two copies of $\CP^1\times \R^+$, so the gauge group is
$(SU(a)\times SU(b)\times SU(c))^2$, with each factor supported
on a separate fixed point component. One set of matter fields we
know about are bi-fundamentals $\Phi_i$ supported on the six
half-lines.
 The ${\Phi}$'s are in the
representations \eqn\judp{\eqalign{&{\bf (a,{\bar b},1,1,1,1) +
(1,b,{\bar c},1,1,1) + (1,1,c,{\bar a},1,1)}\cr &{\bf + (1,1,1,
a,{\bar b},1) + (1,1,1,1,b,{\bar c}) + ({\bar a},1,1,1, 1,
c)}.\cr}} We conjecture that there  are also three chiral
multiplets ${\psi}_m$ which live only at the conical singularity
and are in the representations ${\bf (a,1,1,{\bar a},1,1)}$, ${\bf
(1,b,1,1,{\bar b},1)}$, and  ${\bf  (1,1,c,1,1,{\bar c})}$. Since
the ${\Phi}$'s live on half-lines, we need to specify boundary
conditions on them at the origin. To make a simple proposal, we
write a hypermultiplet $\Phi$ as a pair of ${\cal N}=1$ chiral
superfields $(u,\bar v)$.  A reasonable conjecture is that the
boundary conditions are such that in the absence of Higgsing,
 the only non-zero components at the origin are ${u}_1$, ${u}_2$,
${u}_3$ and $v_4$, $v_5$, $v_6$. The idea is then that when the
$\psi$'s are nonzero (which is supposed to correspond to Higgsing
to the orbifold $X'$) the boundary conditions are deformed to
$u_4\sim \psi_1u_1$ and similarly for the others. It is plausible
that the resulting spectrum could agree with $M$-theory on the
smoothed out cone. However, in order for us to make a precise
proposal, we really have to understand the superpotential of the
model and the boundary conditions after Higgsing.

\subsec{More General Singularities}

Obviously, these examples have many generalizations, obtained as
$U(1)$ quotients of more general conical hyperkahler manifolds.

A simple class of examples are ``toric'' hyper-Kahler manifolds,
with $H=U(1)^k$ acting on $k+2$ hypermultiplets.  (Of course, one
would also like to study examples with nonabelian $H$.)  The
charges are now given by a $k$ $\times$ $(k+2)$ matrix. The
orbifold singularities in this case are encoded in a $(k+2)$-gon,
generalizing the triangle. In these examples the basic nature of
the physics is qualitatively similar to the cases with $k$ $=1$.
The main difference is that one finds more and more gauge groups
and matter fields present. For instance, the symmetries coming
from the $C$-field in the cone on $Y$ are $U(1)^{k+1}$ which gets
broken to $U(1)^k$ when we smooth out the conical singularity.

It is also interesting to consider the toric hyper-Kahler
manifolds as inputs for the construction considered in section
two.  Let $\hat X=\H^{k+2}//H$. Since a set of $k+2$
hypermultiplets admits an action of $U(1)^{k+2}$ which contains
and commutes with $H$, $\hat X$ admits a symmetry group  $F\cong
U(1)\times U(1)$ that preserves the hyper-Kahler structure. The
moment map of $F$ has six components (three for each $U(1)$) so it
gives a natural map $\hat X\to \R^6$ which commutes with the
action of $F$.

Now suppose we pick a $U(1)$ subgroup of $F$ and let $X=\hat
X/U(1)$.  In the spirit of section 2, we may hope that $X$ admits
a conical $G_2$ metric, though this remains to be established. In
any event, $X$ admits the action of $U(1)'=F/U(1)$ and the moment
map on $\hat X$ induces a $U(1)'$-invariant map $X\to \R^6$.
Using the $U(1)'$ orbits to induce a Type IIA description of
$M$-theory on $\R^4\times X$, we find that this model is
equivalent to Type IIA on $\R^4\times \R^6=\R^{10}$, with
$D6$-branes coming from $U(1)'$ fixed points.  Using the
properties of the hyper-Kahler moment map, one can show that the
$D$-brane worldvolumes are all copies of $\R^4\times \R^3$,
linearly embedded in $\R^4\times \R^6$. This conjecturally
generalizes the situation considered in section 3 of \aw.

Since the examples we have studied here are much more explicit
than those in section 2, one might wonder if the examples in
section 2 can be obtained as twistor spaces.  This appears to be
untrue, since for example the weighted projective space
$\WCP^3_{n,n,m,m}$ discussed in \aw\ and in section 2 appears not
be be a twistor space.

\bigskip
Work of EW was supported in part by NSF Grant PHY-0070928. We
would like to thank C. Boyer, K. Galicki, D. Morrison, and S.
Salamon for discussions. \listrefs
\end